\documentclass[pre,twocolumn,showpacs]{revtex4}
\usepackage{amsmath,psfig}
\newcommand{\beq}{\begin{equation}}
\newcommand{\eeq}{\end{equation}}
\newcommand{\bea}{\begin{eqnarray}}
\newcommand{\eea}{\end{eqnarray}}
\newcommand{\bean}{\begin{eqnarray*}}
\newcommand{\eean}{\end{eqnarray*}}
\newcommand{\ba}{\begin{array}}
\newcommand{\ea}{\end{array}}
\newcommand{\bml}{\begin{mathletters}}
\newcommand{\eml}{\end{mathletters}}
\newcommand{\eacc}{{E}_{\rm acc}}
\newcommand{\ecr}{{E}_{\rm cr}}
\newcommand{\rem}[1]{ }
\newcommand{\ga}{\gtrsim}
\newcommand{\la}{\lesssim}
\begin{document}
\title{A Constraint on Electromagnetic Acceleration of Highest 
Energy Cosmic Rays}
\author{Mikhail V. \surname{Medvedev} }
\thanks{ Also at the Institute for Nuclear Fusion, 
Russian Research Center ``Kurchatov Institute", Moscow 123182, Russia}
\email{medvedev@ku.edu}
\affiliation{Department of Physics and Astronomy, University of Kansas,
Lawrence, KS 66045}

\begin{abstract}
The energetics of electromagnetic acceleration of ultra-high-energy 
cosmic rays (UHECRs) is constrained both by confinement of a
particle within an acceleration site and by radiative energy
losses of the particle in the confining magnetic fields.
We demonstrate that the detection of $\sim3\times10^{20}$~eV events 
is inconsistent with the hypothesis that compact cosmic accelerators 
with high magnetic fields can be the sources of UHECRs. This rules 
out the most popular candidates, namely spinning neutron stars, 
active galactic nuclei (AGNs), and $\gamma$-ray burst blast waves. 
Galaxy clusters and, perhaps, AGN radio lobes remain the only 
possible (although not very strong) candidates for UHECR acceleration 
sites. Our analysis places no limit on linear accelerators.
With the data from the future {\it Auger} experiment one should 
be able to answer whether a conventional theory works or some new 
physics is required to explain the origin of UHECRs.
\end{abstract}
\pacs{41.60.-m, 96.40.-z}
\maketitle

\section{Introduction}
The detection of ultra-high-energy cosmic rays (UHECRs) with 
energies above $10^{20}$~eV has posed a challenge to the 
understanding of their origin and nature. At present, 17 such 
events were reported by the {\it AGASA} group and 2 events 
were observed by {\it HiRes} 
an energy of the one of them is estimated to be 
$\sim3\times10^{20}$~eV, close to the energy of the largest 
($\sim3.2\times10^{20}$~eV) event observed with the {\it Fly's Eye}
detector 
Nearly isotropic distribution on the
sky and the absence of large-scale clustering 
suggests the cosmological origin of UHECRs. This is in apparent 
conflict with the observed energy spectrum which lacks the 
Greisen-Zatsepin-Kuzmin (GZK) cutoff at energy 
$\sim5\times10^{19}$~eV \cite{G66,ZK66} 
indicating that UHECRs have traveled the distance smaller than $\sim60$~Mpc
($\sim20$~Mpc for protons with $E\ga3\times10^{20}$~eV).

Cosmic rays are accelerated in astrophysical sources either by
repeated scattering off macroscopic flows, such as shocks, 
winds and outflows, turbulent flows, or directly by an 
induced electric field around a magnetized rotating object.
For the acceleration to operate, a particle must remain
confined within the acceleration region: the gyro-radius of 
the particle should not exceed the size of the system, $R$.
This sets the maximum energy of the accelerated particle,
\beq
\eacc= Ze\,B\,R
\simeq9.3\times10^{23}\,Z\,B\,R_{\rm kpc}\ ~\textrm{eV},
\label{eacc}
\eeq
where $Ze$ is the charge of a particle, $B$ is the 
characteristic magnetic field strength in the acceleration
region (in gauss), and $R_{\rm kpc}$ is the size in kiloparsecs.
If the whole medium is moving relativistically with the Lorentz
factor $\Gamma$ towards an observer, $\eacc$ is boosted to
$\Gamma\eacc$, with $B$ and $R$ being measured in the co-moving frame.
Specifying the particle energy, say $\eacc=3\times10^{20}$~eV
(the largest observed), and assuming $Z\sim1$, one obtains the 
magnetic field--size relation \cite{SB50,Hillas84}, which is plotted 
in Figure \ref{f:1} with the long-dashed line. Any astrophysical 
object to the right from this line can accelerate protons to 
energies $\ge3\times10^{20}$~eV.

Various energy losses, such as collisional and inverse Compton
scattering in a radiation field, limit the UHECR energy.
But even in the absence of any such processes, an energetic 
particle will still loose energy radiatively while moving
through magnetic and electric fields which confine and
accelerate this particle. The radiative (synchrotron) 
losses have been considered by various authors 
\cite{
Hillas84,KE86,Aharonian+02}, who usually
estimated the synchrotron cooling time, but didn't take into 
account the finite size of the source self-consistently.\footnote{
The paper by \cite{Aharonian+02}, appeared when the present paper
was in preparation, accounts for the finite source size but still 
not self-consistently.}

In this paper we re-analyze the energetics of electromagnetic 
acceleration to elucidate its inherent and inevitable limitations. 
The Hillas criterion, equation (\ref{eacc}), is shown to be often
inaccurate and even misleading.
We consider the most idealized models that carry
only the most robust properties of cosmic accelerators, any details 
of the electromagnetic acceleration process are not important for us. 
Thanks to the simplicity of the models, analytical solutions for
the particle energy, which account for a source size 
self-consistently, were obtained. These results represent the 
most relaxed (and, hence, unavoidable) constraints on the maximum
energy of an electromagnetically accelerated particle.
This implies that the most favorable sources of UHECRs,
such as neutron stars (NSs), active galactic nuclei (AGNs) 
and gamma-ray bursts (GRBs), cannot accelerate protons to the energy 
$\sim3\times10^{20}$~eV and, hence, must be ruled out. 
However, our critetion does not apply to linear accelerators 
(e.g., axial jets) \cite{L76}.
The {\it Auger} experiment may be capable of ruling out the 
remaining candidates, radio lobes of AGNs and galaxy clusters, 
and if it does, the whole conventional astrophysical picture of 
UHECRs as accelerated particles must be revisited.

\section{Inefficient acceleration}
Let us consider diffusive acceleration first. This type of 
acceleration operates in shocks of $\gamma$-ray bursters, 
galaxy clusters, jets from AGNs interacting with the intergalactic 
medium and producing radio lobes, and, perhaps, in AGN cores near 
the base of a jet. Let us consider a shock propagating through
a magnetized medium (inter-cluster medium [ICM], for instance), 
as shown in Figure \ref{f:2}a. Magnetic field may be inhomogeneous 
on a scale $\sim R$. An accelerated particle gains energy 
by repeated scattering off a shock or a flow. After every scattering, 
the particle travels a great distance along the Larmor orbit until 
it returns and gets another kick. As long as the particle moves 
freely in the magnetic field it radiates and slows down. 
We will see that the maximum terminal energy of a particle 
(i.e., when the particle escapes the system), $E$, is determined 
by these radiative losses, but is insensitive to how large the energy, 
$E_0$, of the particle at the shock front is. Therefore, we refer to this
regime as ``inefficient acceleration''.

Let us consider a particle with some initial energy $E_0$ propagating
through a region of a size $R$ filled with a magnetic field $B$.
The energy of the particle gradually decreases according to the 
equation (see \cite{Pomeranchuk40}):
\beq
\frac{d{E}}{dx}=F_{\rm Rad}
=-\frac{2}{3}\left(\frac{Ze}{Am_pc^2}\right)^4\,B^2(x)\,{ E}^2,
\label{dedx}
\eeq
where $x$ is the distance along the particle trajectory,  
$F_{\rm Rad}$ denotes the radiation friction force, and
$Am_p$ is the particle mass. In (\ref{dedx}) only the transverse field 
component enters; we assimed that $B_\perp\sim B$. 
The solution of this equation is
\beq
{E}^{-1}={E}_0^{-1}+\ecr^{-1},
\label{e}
\eeq
where ${E}_0$ is the initial energy of the particle and
\bea
\ecr&=&\frac{3}{2}\left(\frac{Am_pc^2}{Ze}\right)^4
\left(\int_0^R B^2(x)\,dx\right)^{-1} \nonumber\\
&\simeq&2.9\times10^{16}\,\frac{(A/Z)^4}{B^2R_{\rm kpc}}\ 
~\textrm{eV}.
\label{ecr}
\eea
For simplicity, it is assumed here that $B(x)\sim constant$ within 
the system. It must be clear now that no matter how energetic the 
particle is (${E}_0\to\infty$), after traveling through a region 
with a magnetic field its energy will not exceed the critical 
energy $\ecr$. Specifying the energy $\ecr=3\times10^{20}$~eV 
and assuming $A\sim1$, one obtains another $B$ vs. $R$ 
constraint, shown in Figure \ref{f:1} with the 
short-dashed line. All astrophysical sources located above 
this line have $\ecr\le3\times10^{20}$~eV and hence 
cannot accelerate UHECRs.

It should be mentioned that the classical expression 
(\ref{dedx}) for the radiation friction force is valid if 
the wavelength of the emitted radiation is larger than the 
``classical radius'' of a charge, 
$c/\omega_B\gg (Ze)^2/(Am_pc^2)$ 
(where $\omega_B=Ze\,B/Am_pc$), that is for the field 
strengths not exceeding $(Am_p)^2c^4/(Ze)^3$ in the rest 
frame of a particle. This yields the condition for the 
particle energy
\beq
{E}\ll\left(\frac{Am_pc^2}{Ze}\right)^3\frac{1}{B}
\simeq1.9\times10^{31}(A/Z)^3B^{-1}\ ~\textrm{eV},
\eeq
which is satisfied for $E\la10^{21}$~eV for practically all 
sources.

\section{Efficient acceleration}
Let us now consider ``cosmic inductors'' where particles
are accelerated by electric fields induced by rapid
rotation of a magnetized object. Acceleration of this type
should occur in neutron star magnetospheres and around
accreting supermassive black holes in the centers of AGNs.
We naturally assume that the magnetic field has a dipolar 
(or a multipolar) structure, hence field lines are bent
on a scale $\sim R$, as shown in Figure \ref{f:2}b.
Because of rapid rotation, there is an induced electric field
$E_{\rm ind}\simeq |{\bf v\times B}|/c$ which accelerates 
a particle. The maximum value of $E_{\rm ind}$ is achieved 
near the light cylinder \footnote{That is at a cylindrical 
radius $R_{lc}$ at which the linear velocity of magnetospheric 
field lines is close to the speed of light, $\Omega R_{lc}\simeq c$,
where $\Omega$ is the angular velocity of a magnetized object,
e.g., a pulsar.} 
is close  where it is equal to, at most, 
$E_{\rm ind}\simeq B$. The particle in such a system gains 
energy rapidly, within one passage through the system. 
Hence one must retain the electromagnetic 
accelerating force $F_{\rm EM}=Ze\,E_{\rm ind}\simeq Ze\,B$ 
in the energy equation (\ref{dedx}). Then it reads
\beq
\frac{d{E}}{dx}\simeq
Ze\,B-\frac{2}{3}\left(\frac{Ze}{Am_pc^2}\right)^4\,
B^2\,{E}^2.
\label{dedx2}
\eeq
For a small initial energy of an accelerated particle 
($E_0\ll{\eacc},{\ecr}$), the solution of this equation 
takes a simple and elegant form:
\beq
E=\sqrt{\eacc\ecr}\;\tanh\!\sqrt{{\eacc}/{\ecr}},
\eeq
where $E$ is the terminal energy of the particle.
The solution has two obvious asymptotics. If $\eacc\ll\ecr$
one recovers the Hillas constrain \cite{Hillas84}
$E\simeq\eacc$ (equation [\ref{eacc}]), whereas in 
the opposite limit one has
\bea
E_{\rm max}&\simeq&\sqrt{\eacc\ecr}\nonumber\\
&\simeq&1.3\times10^{20}A^2Z^{-3/2}B^{-1/2}\ ~\textrm{eV}.
\label{f=f}
\eea
This equation, in fact, follows from the balance between the
acceleration and radiative losses, $F_{\rm EM}=F_{\rm Rad}$,
in equation (\ref{dedx2}). Hence, we refer to this regime
as ``efficient acceleration''.
Acceleration above this energy is impossible because at larger 
energies radiative friction begins to dominate over 
electromagnetic acceleration. This constraint for $E_{\rm max}=
3\times10^{20}$~eV is plotted in Figure \ref{f:1} with the 
dot-dashed line. No objects above this line can accelerate 
cosmic rays to this energy.

\section{Discussion}
The conventional astronomical picture for the origin of these 
cosmic rays is the acceleration of charged particles, e.g., 
protons or heavier atomic nuclei, in extragalactic objects
\cite{Hillas84,Hillas98}. There are only few types of such 
objects. The most favorable are: spinning neutron stars (NSs) 
and magnetars 
central regions of active galactic nuclei (AGNs) 
AGN radio lobes 
and $\gamma$-ray burst (GRB) shocks. 
It is unlikely that shocks in galaxy 
clusters are the UHECR sources because they are too far, beyond 
the GZK distance, and they are not able to accelerate protons to 
energies above few times $10^{19}$~eV \cite{KRJ96}.

\begin{figure}
\psfig{file=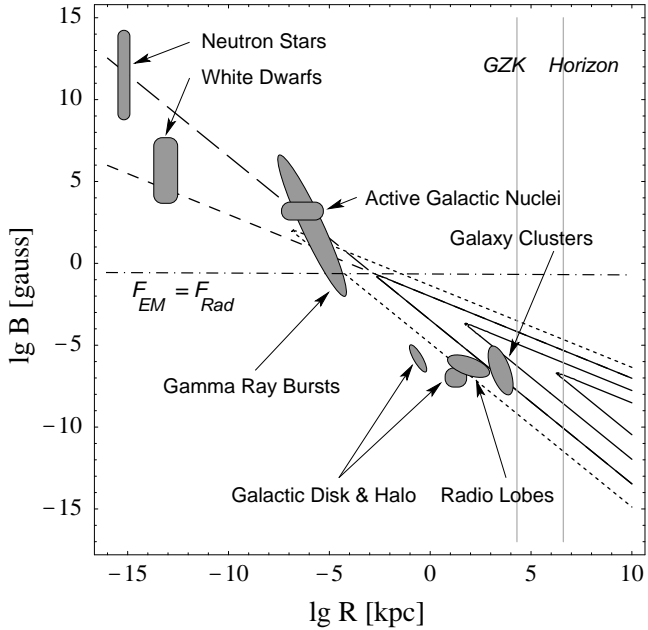} 
\caption{The $B$ vs. $R$ diagram for UHECR sources.
The long-dashed line is the original Hillas relation, eqn. (\ref{eacc}), 
for a proton of energy $3\times10^{20}$~eV. The short-dashed and 
dot-dashed lines represent the radiative cooling constraints for the 
diffusive and inductive acceleration, given by eqns. (\ref{ecr}) and 
(\ref{f=f}), for the same proton energy. Note that above the dot-dashed 
line the force of radiative friction dominates over any electromagnetic 
forces. The dotted and solid lines represent the boundaries of the 
allowed parameter regions for $3\times10^{20}$~eV iron nuclei and
for protons of energies $3\times10^{20}$, $10^{22}$, and 
$3\times10^{23}$~eV, respectively. Only those astronomical objects
which fall inside the ``wedges'' are, in principle, capable of 
accelerating the particles to such energies. The gray vertical
lines mark two characteristic scales: the GZK attenuation distance 
($\sim 20$~Mpc) and the Hubble horizon size ($\sim 4$~Gpc).
For GRBs we took into account that the Lorentz boost changes with 
radius.
\label{f:1} }
\end{figure}

\begin{figure}
\psfig{file=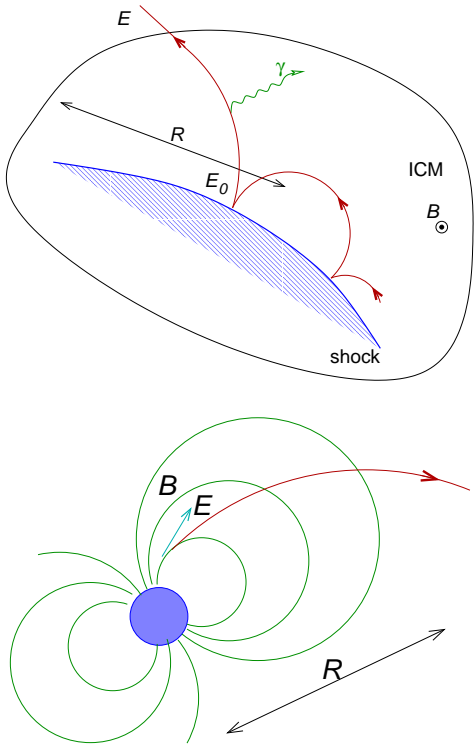}
\caption{Cartoons representing 
a typical system with inefficient (diffusive) acceleration (a)
and with efficient (inductive) acceleration (b).
\label{f:2} }
\end{figure}

In the previous sections we demonstrated that the energy 
of a particle confined by magnetic fields within an 
acceleration site is determined by its radiative losses.
We now discuss the constraints on $B$ and $R$ given by 
equations (\ref{eacc}), (\ref{ecr}), and (\ref{f=f}) and
plotted in Figure \ref{f:1}.

First, the constraint (\ref{f=f}) shown with the dot-dashed 
line is the most stringent. It tells that protons cannot be 
accelerated electromagnetically at the sources which lie 
above this line. 
Equation (\ref{f=f}) also holds for heavier (e.g., iron) 
nuclei. Hence, compact stars, AGN cores, and GRB shocks 
(except during the late afterglow phase) are readily ruled 
out from the list of possible sources of ultra-high-energy 
protons and nuclei.

Second, the remaining candidates, i.e., the radio lobes
and galaxy clusters, may accelerate particles only via the 
diffusive mechanism, hence equation (\ref{ecr}) is 
appropriate. This equation together with (\ref{eacc})
specifies the allowed $B$--$R$ parameter region. In the figure
the dotted line corresponds to iron nuclei with the energy
$3\times10^{20}$~eV and the solid lines correspond to protons  
with three energies: $3\times10^{20}$, $10^{22}$, and 
$3\times10^{23}$~eV. One can see that the AGN radio lobes 
are at most marginally consistent with being the sources 
of the highest energy cosmic rays. Moreover, only a handful 
of such sources are relatively nearby (e.g., M87, Cen~A, 
NCG~315) but their angular distribution is completely 
uncorrelated with the nearly isotropic distribution arrival 
directions of UHECRs. 
Shocks in galaxy 
clusters can be such sources, according to Figure \ref{f:1}.
It has been argued from a more detailed analysis of shock
acceleration, however, that they are not able to accelerate 
protons above $\sim6\times10^{19}$~eV \cite{KRJ96}.
Overall, large objects are more preferable candidates for the
highest energy cosmic ray sources, both due to the larger
terminal energy of an accelerated particle and the larger
energy reservoir available, see Fig. 5 in the paper by
Kronberg \cite{K02}.  

Equation (\ref{eacc}) together with (\ref{ecr}) or (\ref{f=f})
puts a lower bound \footnote{The above criteria are applicable 
only to cosmic accelerators which involve acceleration on curved 
paths; no constraints on linear cosmic accelerators (if any) 
are placed.} on the size of the accelerating source:
\beq
R\ga 6.0\times10^{-5}\,E_{20}^3\,Z^2\,A^{-4}~ \textrm{ kpc},
\label{r}
\eeq
where $E_{20}=E/10^{20}$~eV. This is a quite remarkable 
result since it sets the absolute limit on $R$, independent
of the field strength. Now, two special cases follow.
First, the size $R$ exceeds the GZK distance $\sim20$~Mpc 
when the energy of the proton is larger than 
$E_{\rm GZK}\sim7\times10^{22}$~eV. That is, such an energetic
proton will loose its energy through the interaction with the
2.7~K background radiation right at the acceleration site.
Hence, it is unlikely that protons may be accelerated to the
energies above $E_{\rm GZK}$. Second, above the energy
$E_{\rm Hor}\sim4\times10^{23}$~eV, the size of the 
accelerator exceeds the size of the Universe. Thus, 
$E_{\rm Hor}$ is the ultimate upper bound on the energy of 
electromagnetically accelerated protons. Note that this 
energy limit is below the energy of primary protons 
($\sim10^{24}$~eV) in the most popular (among other models 
\cite{Sarkar02,Watson01}) Z-burst model.


To conclude, we arrived at an interesting result. 
Practically all known astronomical sources are not able 
to produce cosmic rays with energies near few times 
$10^{20}$~eV. There is not too much room left for the 
conventional electromagnetic (in a broad sense) acceleration.
Pushing observations up in energy by about an order 
of magnitude should clarify this situation. During ten 
years of operation of the {\it Auger} cosmic ray 
observatory
about 300 events above $10^{20}$~eV are 
expected to be recorded. Extrapolating the present 
{\it AGASA} spectrum one can then expect a few events
near or above $10^{21}$~eV to be recorded, which may be
enough to exclude radio lobes from UHECR sources. 
Should this happen, a considerable revision of the 
current astrophysical picture will be inevitable.

\section*{Acknowledgements} 
We thank Andrei Beloborodov for a discussion and
an anonymous referee for very careful reading 
and many suggestions greatly improving the manuscript.

\bigskip

\vfill



\end{document}